\documentclass[review]{elsarticle}
\graphicspath{ {./figures/} }
\usepackage{hyperref}
\usepackage{float}
\usepackage{verbatim} 
\usepackage{apalike}
\usepackage{amsmath,amsfonts,amssymb}
\restylefloat{figure}
\restylefloat{table}
\usepackage{xcolor}
\hypersetup{
    colorlinks,
    linkcolor={red!50!black},
    citecolor={blue!50!black},
    urlcolor={blue!80!black}
}
\journal{Expert Systems with Applications}

\biboptions{authoryear}

\begin{document}
\begin{frontmatter}

\begin{titlepage}
\begin{center}
\vspace*{1cm}

\textbf{A Novel Experts' Advice Aggregation Framework Using Deep Reinforcement Learning for Portfolio Management}

\vspace{1.5cm}

MohammadAmin Fazli$^a*$ (fazli@sharif.edu), Mahdi Lashkari$^a$ (mahdi.lashkari@sharif.edu), Hamed Taherkhani$^a$ (hamed.taherkhani@sharif.edu), Jafar Habibi$^a$ (jhabibi@sharif.edu) \\

\hspace{10pt}

\begin{flushleft}
\small  
$^a$ Department of Computer Engineering, Sharif University of Technology, Tehran, Iran \\

\vspace{1cm}
\textbf{Corresponding Author:} \\
MohammadAmin Fazli \\
Department of Computer Engineering, Sharif University of Technology, Tehran, Iran \\
Tel: (+98) 937-169-2023 \\
Email: fazli@sharif.edu

\end{flushleft}        
\end{center}
\end{titlepage}

\title{A Novel Experts' Advice Aggregation Framework Using Deep Reinforcement Learning for Portfolio Management}

\author[label1]{MohammadAmin Fazli \corref{cor1}}
\ead{fazli@sharif.edu}

\author[label1]{Mehdi Lashkari}
\ead{lashkari.ma@gmail.com}

\author[label1]{Hamed Taherkhani}
\ead{hamed.taherkhani@sharif.edu}

\author[label1]{Jafar Habibi}
\ead{jhabibi@sharif.edu}
\cortext[cor1]{Corresponding author.}
\address[label1]{Department of Computer Engineering, Sharif University of Technology, Tehran, Iran}

\begin{abstract}
Solving portfolio management problems using deep reinforcement learning has been getting much attention in finance for a few years. We have proposed a new method using experts' signals and historical price data to feed into our reinforcement learning framework. Although experts' signals have been used in previous works in the field of finance, as far as we know, it is the first time this method, in tandem with deep RL, is used to solve the financial portfolio management problem. Our proposed framework consists of a convolutional network for aggregating signals, another convolutional network for historical price data, and a vanilla network. We used the Proximal Policy Optimization algorithm as the agent to process the reward and take action in the environment. The results suggested that, on average, our framework could gain 90 percent of the profit earned by the best expert.
\end{abstract}


\end{frontmatter}

\section{Introduction}
\label{introduction}
Stock trading is about buying and selling stocks in the financial market to maximize the profit earned from the investment—profit-making can be achieved by exploiting price fluctuations in the market. When an investor sells stocks at a higher price, a profit is made. However, the price fluctuations are usually so high and the environment so dynamic that they make trading optimization hard to achieve. As a result, the situation in the stock market obliges investors to use an intelligent policy that can adapt itself to the dynamic and non-deterministic idiosyncrasies of stock markets. 
\par Portfolio management is the decision-making process of continuously reallocating an amount of found into different financial investment products, aiming to maximize the return while restraining the risk \cite{haugen1988}. Portfolio management systems often use price history fluctuations of one or multiple stocks. However, besides the price history, there is other information in the stock markets to take advantage of, including textual, macroeconomics, knowledge graph, image, fundamental, and analytics data \cite{jiang2021}. Many models, such as ANN, DNN, PCA, RNN, CNN, LSTM, attention models, and others, have been used in recent studies to address the stock market prediction problem \cite{jiang2021}. These models have been used in multiple markets, including the US, China, Japan, Korea, India, Brazil, and others. Some works regarding financial time series forecasting, such as stock price, index, commodity price, volatility, bond price, forex price, cryptocurrency, and trend forecasting using deep models, have been done \cite{sezer2020}.
\par The advent of reinforcement learning (RL) in financial markets is driven by several advantages inherent to this field of artificial intelligence. In particular, RL allows the combination of the "prediction" and the "portfolio construction" task in one integrated step, thereby closely aligning the machine learning problem with the objectives of the investor \cite{fischer2018reinforcement}. Three main approaches of deep RL in finance have been elaborated and compared, indicating that the actor-only approach is the best-suited one for financial markets \cite{fischer2018reinforcement}. Many types of research have been made to use RL, and deep RL in finance \cite{neuneier1997enhancing, cumming2015investigation, watts2015hedging, du2016algorithm, charpentier2021reinforcement, taghian2022learning}. Some other efforts are made to address the portfolio management problem using RL \cite{jiang2017, liang2018adversarial, jeong2019improving, filos2019reinforcement, ye2020reinforcement, aboussalah2020continuous}. Solving portfolio management problem using modern deep reinforcement learning approaches constitutes only a small part of problems addressed in stock markets. Hence side resources have yet to be fully taken advantage of in these frameworks. An essential source of information in the field of finance is the experts' signals. Some research has been made to exploit experts' advice, and signals in the field of finance, especially for portfolio management \cite{hill2011, bhattacharya2013, he2021, zhang2017, yang2020, yang2020boosting, wu2020}. We will thoroughly investigate them in the next section.
\par In this paper, alongside the price history, we used signals from experts to feed into our deep RL network. Due to the dynamic and not-deterministic features of the stock markets and the fact that the profit is determined in the long and mid-term, we enjoyed a deep reinforcement learning framework to model the problem. In this framework, using a reward function, we tried to maximize the accumulated profit in the long term; that is, our goal is to achieve a policy that may not necessarily be profitable in the short or mid-term, but it is aimed to be optimal in the long term.

\section{Related Works}
\label{Related}
In the field of experts' opinion aggregation, to achieve a fair and accepted decision, it is a good practice to average the opinions of experts who have access to different resources. We can leverage these signals to help get an accurate prediction of stocks' future returns.
\par A crowd opinion aggregation model, namely CrowdIQ, has been proposed, which has a differential weighting mechanism and accounts for individual dependence \cite{du2017}. A decay function was introduced to address the dependency problem in individual judgments and give different weights to judges based on their previous judgments. Those judges who make original judgments were given higher weights than those who follow others' judgments. CrowdIQ has been evaluated compared to four baseline methods using real data collected from StockTwits. The results showed that CrowdIQ significantly outperformed all baseline methods in terms of both a quadratic prediction scoring method and simulated investment returns. The research sample is relatively small, and more stocks and postings are needed to improve the generalizability of the results.
\par A genetic algorithm approach that can be used to identify the appropriate vote weights for users based on their prior individual voting success has been proposed  \cite{hill2011}. These are user-generated stock pick votes from a widely used online financial newsletter. This method allowed them to identify and rank “experts” within the crowd, enabling better stock pick decisions than the S\&P 500. They showed that the online crowd outperformed the S\&P 500 in two test periods, 2008 and 2009. The main disadvantage is that the period covered by the data set was too short to be ideal for testing and training. Also, during the testing period, the market was abnormal, meaning that the result was not indicative of future performance.
\par A persuasion game between a set of experts and a judge is applied to find the efficiency of decision-making \cite{bhattacharya2013}. Each expert is identified by her quality and agenda, meaning they can be biased. This article tries to find the relation between \emph{the conflict of experts} and the quality of decision-making. The finding suggests that, first, employing better quality experts does not necessarily lead to better decision-making; second, it is optimal for the judge to choose experts whose agendas are either extremely opposite or aligned; and third, it may be optimal to employ two experts with the same extreme agenda rather than experts with completely opposite ones.
\par An experiment was made to answer the question of whether the wisdom of crowds helps or hinders knowledge of factual information. \cite{he2021} The results show that policymakers tend to rely heavily on majority rule as the preferred heuristic, and also, there is a limit to the amount of information about the opinions of others that they can effectively process. In addition, information about others’ answers improves performance on easy questions but tends to harm performance on difficult ones, showing the downside of this work. 
\par Conventionally, there are two main approaches to stock portfolio selection, the mean-variance model and capital growth theory. The variance model focuses on creating a balance between return value and risk in the portfolio of a period. However, CGT focuses on maximizing the expected growth rate of the portfolio or the expected logarithmic return of the multi-period or sequential portfolio. The online selection of stocks is a problem in which several selection periods are considered, and a statistical hypothesis on the future prices of assets is not determined. The weak aggregation algorithm (WAA) \cite{kalnishkan2008} is an online sequencing algorithm that makes decisions using the weighted average of all expert recommendations and updates each expert's weight based on their previous performance.
\par Based on the WAA algorithm, a method in which expert opinion is used for learning and decision-making for the online stock portfolio selection problem was introduced \cite{zhang2017}. This method assigns a weight to each expert based on their performance in previous rounds. This algorithm is defined in two ways: WAAS, in which an expert recommends a stock, and WAAC, in which an expert recommends a stock portfolio containing multiple assets. The two defined algorithms are designed to work without considering information from the financial markets. A new algorithm is defined in which the WAAC is combined with the ancillary information of the financial markets and makes a new strategy called WAACS \cite{yang2020}. In other researches, using the WAA method to collect expert recommendations, a new online portfolio selection strategy called continuous aggregating exponential gradient (CAEG) was proposed \cite{ yang2020boosting, he2020universal, yang2022}. In this method, first, a set of exponential gradient strategies with different learning rates g are considered experts, and then the stock portfolio is calculated in the next period using the WAA method to aggregate all the recommendations.
\par A novel arbitrage-based approach in which multiple prediction models are dynamically combined to obtain predictions is presented \cite{cerqueira2019}. Arbitrage is a meta-learning approach that combines the output of experts based on predictions of the loss they will suffer. It is argued that this meta-learning approach is useful to better capture recurring changes in the environment. This proposal also includes a sequential rebalancing approach to model dependencies between experts. \cite{mcandrew2021} reviewed the current literature on the aggregation of predictions obtained from peers. They have compiled common terminology, aggregation methods, and forecast performance metrics. They identified four key themes that are repeated in the combined forecasting literature: (a) using human intuition to facilitate statistical forecasting when data is sparse and changing rapidly, (b) involving experts because of their role as decision-makers, (c) Using more straightforward aggregation of models to combine prediction densities and more complicated models to combine point predictions and (d) lack of experimental design and comparative measurements in many manuscripts.
\par According to the positive impact of two sources of stock analysts' opinions and information from collective wisdom (social network users), \cite{eickhoff2016} examine the relationship and its impact on predictions. The important issue is that stock analysts, individuals who are paid to provide regularly updated opinions about specific assets, may face restrictions that influence their opinions, whereas such influence would not be problematic at the level of social media users. Results suggest that there is no uniform orientation between these two areas, professional analysts can include information before social media in their evaluation, and if the reasons for changes are known to the public, social networks can change faster. Different effects of analyst attitude and crowd sentiment on stock prices are compared \cite{wu2020}. They found that the wisdom of the experts and the crowd are positively related to stock prices in the following days. They adopted LightGBM \cite{ke2017lightgbm} to predict the trend of stocks, suggesting that the wisdom of the crowd is more valuable in investing than the wisdom of experts.
\par A tool, named SINVLIO, was presented based on semantic technologies and fuzzy logic techniques that recommend investments based on both the psychological aspects of the investor and the traditional financial parameters of investments \cite{garcia2012sinvlio}. Jörg Gottschlich and Oliver Hinz proposed a decision support system design that allows investors to include crowd recommendations in their investment decisions and use them to manage a portfolio \cite{gottschlich2014}.  In \cite{li2010}, fuzzy and genetic algorithms were used to solve the financial portfolio management problem. The planning agent can easily access all the intelligent processing agents, including the financial risk tolerance evaluation, asset allocation, stock portfolio selection, interest prediction, and decision aggregation agents. However, in this method, a suitable mechanism is not defined for selecting specialists. It is possible to use various sources of collective wisdom in investing management. Kumar et al. used the potential of virtual investing communities(VIC) such as MotleyFools CAPS, Sharewise, Minkabu, Hollywood Stock Exchange, etc \cite{kumar2020deriving}. The idea of this research is the automatic construction of a stock portfolio that is dynamically adjusted and relies on VIC information. In order to combine expert domain knowledge with collective wisdom, Kovacevich presented a framework that enables ranking and selection of alternatives and quantifying the quality of crowd votes \cite{kovacevic2020crex}. This framework permits the consideration of the crowd's votes concerning the knowledge of the experts and the modeling procedures of the compromises between the crowd and the satisfaction of the experts in the final decisions.
\section{Preliminaries}
\label{Preliminaries}
We intend to design a model for the portfolio management problem based on experts' advice aggregation. In this model, an intelligent agent finds major stock market patterns and takes advantage of them in tandem with signals generated by experts to optimize the portfolio selection problem.
\subsection{Mathematical Prerequisites}
Inspired by \cite{jiang2017}, a mathematical model of an RL-based portfolio management problem is represented in this section. This problem is divided into steps in which the agent makes a decision based on the environment. That is, the time is segmented into equal parts. In every episode, the agent distributes the capital into different stocks. These episodes continue until they reach a final point. It is important to mention that in this article, we assumed that every episode is a day long.
The portfolio can hold m assets. The closing prices of all assets include the price vector for period t, $v_t$. The relative price vector of the t-th trading period, $y_t$, is defined as the item-by-item division of $v_t$ by $v_{t-1}$:
\begin{displaymath}
y_t := v_t \varnothing v_{t-1} = (1,\dfrac{v_{1,t}}{v_{1,t-1}},
\dfrac{v_{2,t}}{v_{2,t-1}},...,
\dfrac{v_{m,t}}{v_{m,t-1}}
)^\intercal.
\end{displaymath}
The elements of $y_t$ are the quotients of the closing prices for each asset during the period. The relative price vector can be used to calculate the change in total portfolio value over a period. If $p_{t-1}$ is the value of the portfolio at the beginning of period $t$, without taking into account the transaction cost,
$$
p_t=p_{t-1} \boldsymbol{y}_t \cdot \boldsymbol{w}_{t-1}
$$
where $w_{t-1}$ is the portfolio weight vector (henceforth called portfolio vector) at the beginning of period t, whose ith element, $w_{t-1,i}$, is the proportion of asset $i$ in the portfolio after reallocation of capital. The rate of return for period $t$ is defined as
$$
\rho_t:=\frac{p_t}{p_{t-1}}-1=\boldsymbol{y}_t \cdot \boldsymbol{w}_{t-1}-1
$$
In a real scenario, buying or selling assets in a market is not free. The cost typically amounts to the commission fee. Reallocation to a new portfolio shrinks the portfolio value by a factor of $\mu_t$. Considering the transaction cost in the formula, we need to rewrite the rate-of-return formula below
$$
\begin{aligned}
&\rho_t=\frac{p_t}{p_{t-1}}-1=\frac{\mu_t p_t^{\prime}}{p_{t-1}}-1=\mu_t \boldsymbol{y}_t \cdot \boldsymbol{w}_{t-1}-1 \\
\end{aligned}
$$
and logarithmic rate of return as
$$
r_t=\ln \frac{p_t}{p_{t-1}}=\ln \left(\mu_t \boldsymbol{y}_t \cdot \boldsymbol{w}_{t-1}\right) \\
$$
Finally, the final portfolio value will be
$$
p_{\mathrm{f}}=p_0 \exp \left(\sum_{t=1}^{t_{\mathrm{f}}+1} r_t\right)=p_0 \prod_{t=1}^{t_{\mathrm{f}}+1} \mu_t \boldsymbol{y}_t \cdot \boldsymbol{w}_{t-1}
$$
where $p_0$ is the initial investment amount. The job of a portfolio manager is to maximize $p_f$ for a given time frame.
The transaction remainder factor($\mu_t$) will be calculated as follow
$$
\mu_t=\frac{1}{1-c_{\mathrm{p}} w_{t, 0}}\left[1-c_{\mathrm{p}} w_{t, 0}^{\prime}-\left(c_{\mathrm{s}}+c_{\mathrm{p}}-c_{\mathrm{s}} c_{\mathrm{p}}\right) \sum_{i=1}^m\left(w_{t, i}^{\prime}-\mu_t w_{t, i}\right)^{+}\right]
$$
where $c_p$ and $c_s$ are the commission rates for purchasing and selling stocks respectively. In this article, we considered the sum of the two commission rates as one-tenth of a percent.
\par
We restricted the total episodes during which our model is training. We defined three rules according which the model stops the ongoing training and then starts a new one.
\begin{enumerate}
    \item Absence of new data(at the end of training data samples)
    \item A $10\%$ decrease in the stock portfolio's value compared to the starting point. The higher this criterion is considered, the higher the risk of the model will be, but it will enable the model to find more forward-looking strategies. This number can be adjusted during training and considered zero during testing.
    \item A $20\%$ decrease in the stock portfolio's value compared to the maximum point of the ongoing episode. This criterion is helpful in a situation where the agent achieves a high profit in the current episode but continues to lose this profit, which should be stopped at some point. The higher value we consider for this criterion, the agent will be able to find more promising strategies, which, of course, will also face a high risk.
\end{enumerate}
\begin{figure}[htbp]
\centerline{\includegraphics[scale=.8]{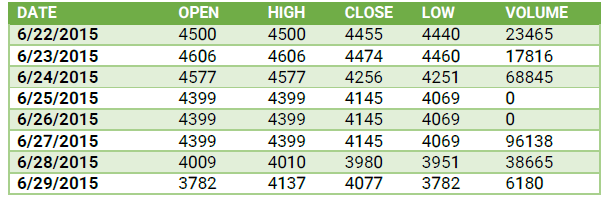}}
\caption{Sample price data of a stock during a week}
\label{fig2}
\end{figure}
\section{Data Treatments}
\label{Data}
In this article, we have leveraged two kinds of information which are elaborated in the following sections.
\subsection{Price History}
We used the price history of stocks in the Iran stock market. They have been gathered by a crawler, implemented by one of the authors, from the securities and exchange organization portal. The crawled data include daily prices from 2017 until 2020. The testing phase covers the last six months, and the rest of the data is used for training. Every row in the dataset consists of open, high, close, and low prices and the total traded volume of a single stock during a single day. Figure \ref{fig2} shows an example of one week's data related to a particular stock. The data relating to 54 stocks have been collected and will be used to train and test the model.
\par
\begin{figure}[htbp]
\centerline{\includegraphics[scale=0.8]{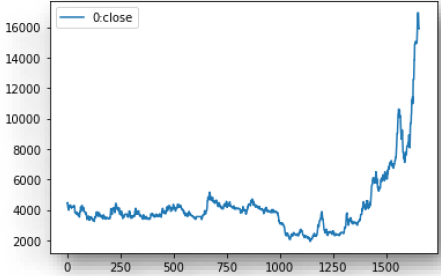}}
\caption{Example of stock price over time}
\label{fig3}
\end{figure}
The statistical and analytical review of the stock price data shows the instability of the data, and as shown in figure \ref{fig3} for one of the stocks, the data has no stability over time, neither in terms of average nor variance.
\par
\begin{figure}[htbp]
\centerline{\includegraphics[scale=0.8]{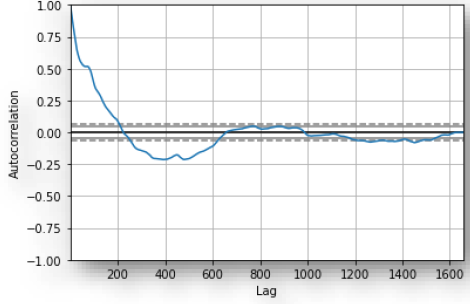}}
\caption{Correlation of a stock's price over time}
\label{fig4}
\end{figure}
In addition, the examination of the correlation criterion of the data shows that the data has zero correlation in a time interval of about 200 days, which will be used in the time frame for training the model. Figure \ref{fig4} shows the correlation of the stock in \ref{fig3}.

\subsection{Experts' Signals}
Experts' signals are obtained from 100tahlil website, containing signals from 85 experts. All the signals of these experts, which are intended for 54 desired stocks, add up to a total of 6335 signals, any of which comprises the following information.
\begin{table}
\begin{tabular}{||c c c c c c||} 
 \hline
 $start date$ & $close date$ & $expected return$ & $expected risk$ & $export$ & $symbol$ \\ [0.5ex] 
 \hline\hline
 6/25/2019 & 7/7/2019 & 48.39 & -14.8 & 49 & 0 \\ 
 \hline
\end{tabular}
\caption{Example of a signal}
\label{table1}
\end{table}
\begin{itemize}
\item Expert Id: the identification of the signal issuer
\item Stock Id: the identification of the stock issued
\item Start date: the beginning date when the signal becomes valid  
\item End date: the date when the signal expires
\item Expected return: the maximum return which may be achieved during start and end date
\item Expected risk: the maximum loss which may occur during start and end date
\end{itemize}
 Table \ref{table1} presents an example signal. Each signal, from the time it is activated until the time it ends, based on the fluctuations of the relevant stock price, will result in profit or loss at any moment. For these calculations not to reduce efficiency during model training, instant profit or loss of each signal is calculated and stored in advance.
\subsection{Managing Missing And Overlapping Data}
During a certain period, some stocks do not have any price on some days because their symbol is closed. To solve this problem, we used the last valid price method, in which whenever there is no price, the nearest next or previous valid price is considered.
\par
 For a particular stock,  Some experts may have provided several signals overlapping over a period of time. We considered the average expected profit and loss in the overlapping periods to deal with this problem.
\begin{figure}[tp]
\centerline{\includegraphics[scale=0.5]{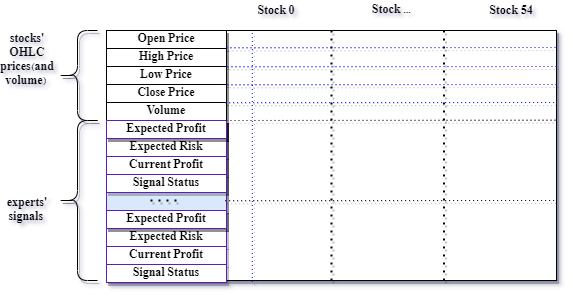}}
\caption{Aggregation of the stocks' OHLC prices and experts' signals}
\label{fig5}
\end{figure}
\subsection{Data Aggregation}
We need to prepare the data to make it usable in our model. Our approach uses the time series in which stocks' signals and price history are divided daily to be fed to the model. For this purpose, we used a time window of 60 days, in which the model will use the stock's historical price and signal during the period. We chose this number based on some experiments. As shown in \ref{fig5}, for every stock, the open, high, low, and close prices and the total trade volume for the stock during a single day are gathered. Every active signal suggested by experts is also added to the aggregated data. Experts' signals include the following information:
\begin{enumerate}
    \item expected return
    \item expected risk
    \item instant return defined as:
$$
\text { InstantReturn }=\frac{\text { CurrenSignalPrice }-\text { StartSignalPrice }}{\text { StartSignalPrice }}
$$
    \item Status: it determines the final state of a signal. A value of zero indicates an active signal. Values 1 and -1 are considered if the signal ended with profit and loss, respectively.

\end{enumerate}
\begin{figure}[tp]
\centerline{\includegraphics[scale=0.4]{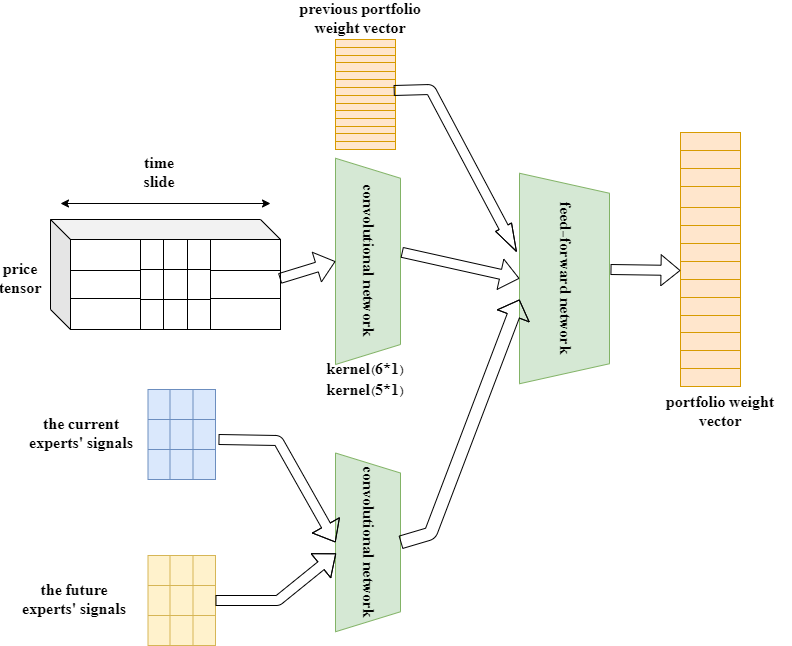}}
\caption{The architecture of proposed method}
\label{fig6}
\end{figure}
\section{Proposed Method}
\label{Method}
In this article, we have proposed a convolutional network as it is a widely used and preferable architecture in the context of the portfolio management problem. In this architecture, 60 days of stock price information (starting price, closing price, low price, high price, and trading volume) are combined in the form of a tensor with 60 days of all experts' signal information and given to the neural network. Every stock is processed independently through two consecutive convolutional layers, using 1*6 and 1*5 kernels, and then is fed to a feed-forward neural network, which uses a tanh activation function to determine the proportion of the stocks as the next portfolio weight vector. As stated before, in the real-world scenario, there are fees to be paid for any transaction, which can add up and negatively impact the profit. In order to prevent constant changes in the portfolio weight vector, the previous portfolio weight vector is added to the output of the convolutional networks and then fed to the feed-forward network. There are two kinds of signals, which are current and future ones. In the current signals, we know the instant profit or loss made by the signal, while in future signals, only the expected profit or loss is known. Figure \ref{fig6} demonstrates the proposed architecture.
\begin{figure}[htbp]
\centerline{\includegraphics[scale=.5]{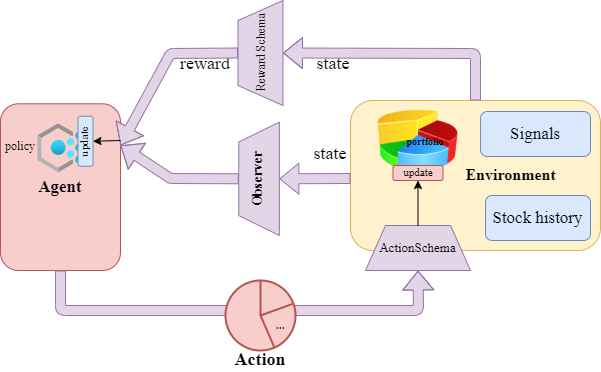}}
\caption{Reinforcement Learning Framework Model }
\label{fig1}
\end{figure}
\section{Experimental Results}
\label{Results}
There are three frameworks, Pytorch, Ray, and Rlib, in use to 
implement the proposed model introduced in the previous section. Rllib is a framework for reinforcement learning modeling and can implement various models. This framework has the capability of distributed execution with the help of Ray. To use the features of Rllib, the environment and the agent we implement must inherit several features from the abstractions of this framework so that different algorithms provided by this framework can be used. For this purpose, we created an environment called TradingEnv, which contains all the necessary facilities to simulate the investment environment and manage the stock portfolio. When the environment receives an action, applying that action to the stock portfolio, it determines the amount of profit earned and provides it to the agent. To simulate the execution of an action in an environment, an ActionSchema, which calculates the change in the stock portfolio's value in every step, is presented. In addition, RewardSchema specifies how to calculate the reward in each step. A general overview of different parts of the reinforcement learning environment is illustrated in figure \ref{fig1}.
\begin{figure}[tp]
\centerline{\includegraphics[scale=.5]{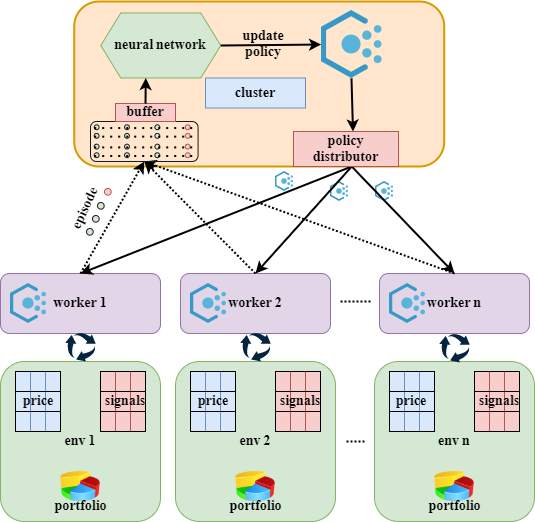}}
\caption{The infrastructure architecture of model execution}
\label{fig7}
\end{figure}
\par
As mentioned, the architecture can also run in a distributed manner. Fig \ref{fig7} describes how it runs distributedly on several servers. Each worker can be a server. In addition, according to the Ray framework's capabilities, it can also use any CPU core as a worker. We can have an arbitrary number of workers, each of which has a copy of the current environment and policy, produces one or more episodes, and then sends them to the leader cluster. Next, the cluster leader updates the original policy and sends this new policy to the workers so that new episodes are generated based on the latest policy.
\par
We chose the PPO(Proxy Policy Optimization) algorithm, considered one of the most effective reinforcement learning algorithms in continuous environments, as the base algorithm for updating our policy. PPO is an Actor-Critic algorithm that creates several episodes of a strategy in each step and then updates the strategy to maximize the average reward. The main advantage of this algorithm compared to other Actor-Critic ones is that in this one, the amount of change is limited in each step, helping to make the variance of changes less and, as a result, lead to more stable models.

\begin{table}[h!]
\begin{tabular}{|c | c|} 
 \hline
 Parameter & The tested value \\ [0.5ex] 
 \hline\hline
 [1e-2, 1e-3, 5e-4, 1e-4, 5e-5, 1e-5] & LEARNING\textunderscore RATE\\ 
 \hline
 [100, 200, 300, 400] & SGD\textunderscore MINIBATCH\textunderscore SIZE\\
 \hline
 uniform(0.85, 1.0) & LAMBDA\\
 \hline
 uniform(0.1, 0.5) & CLIP\textunderscore PARAM\\
 \hline
 [14, 30, 60, 90] & ROLLOUT\textunderscore FRAGMENT\textunderscore LENGTH\\
 \hline
 -0.2 & MAX\textunderscore DRAWDOWN\\
 \hline
 -0.1 & MIN\textunderscore PROFIT\\
 \hline
\end{tabular}
\caption{Examination space for model parameters}
\label{table2}
\end{table}

\begin{table}[h!]
\begin{tabular}{|c | c|} 
 \hline
 Parameter & The selected value \\ [0.5ex] 
 \hline\hline
 5e-4 & LEARNING\textunderscore RATE\\ 
 \hline
 300 & SGD\textunderscore MINIBATCH\textunderscore SIZE\\
 \hline
 0.9 & LAMBDA\\
 \hline
 0.2 & CLIP\textunderscore PARAM\\
 \hline
 60 & ROLLOUT\textunderscore FRAGMENT\textunderscore LENGTH\\
 \hline
 -0.2 & MAX\textunderscore DRAWDOWN\\
 \hline
 -0.1 & MIN\textunderscore PROFIT\\
 \hline
\end{tabular}
\caption{Selected parameters for model training}
\label{table3}
\end{table}

\par
A number of hyperparameters have been tested in this paper. Table \ref{table2} shows the parameters and their tested values. After experimenting with the testing values, we found the best ones, shown in table \ref{table3}.

\begin{figure}[tp]
\centerline{\includegraphics[scale=0.7]{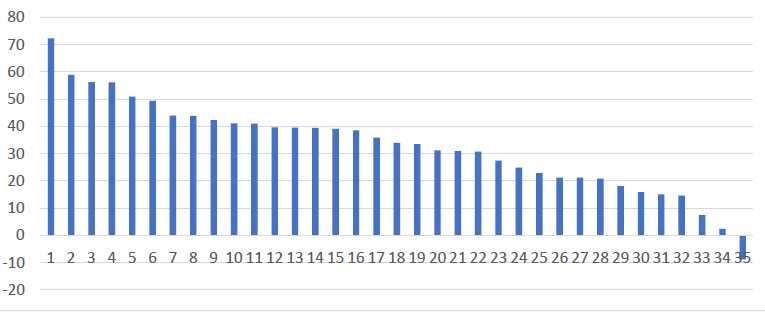}}
\caption{The average profit of each expert's signals during the test period}
\label{fig8}
\end{figure}
We compared the results of our model with the average profit gained by the experts' signals. The average profit is illustrated in figure \ref{fig8}. As can be seen, the best expert averaged 72\% profit, and the worst expert had about 8\% loss during the test period. In addition, the highest profit gained by the best expert was 110\%.

\begin{table}[h!]
\begin{tabular}{|c | c | c | c |} 
 \hline
 Average gain & Maximum gain & minimum gain \\ [0.5ex] 
 \hline\hline
 65\% & 85\% & 32\% \\
 \hline
\end{tabular}
\caption{Average, maximum and minimum gain during all periods}
\label{table4}
\end{table}

The results suggest that, on average, our model is able to earn 90\% of the profit that the best expert earned. Also, in the maximum mode, our model was able to obtain 77\% of the signals obtained by the best expert.

\section{Conclusion}
\label{Conclusion}
This paper proposes a new deep RL framework for the portfolio management problem using Iran's stock market data. We leveraged experts' signals, historical price data, and a deep RL network, which is a new approach to solving this problem, as far as we know. We took advantage of convolutional networks to extract meaning from historical prices and experts' signals and then aggregated them in a feed-forward network to find the next portfolio weight vector. Despite being simple, we found that our model performed very well and could compete with the best experts. We showed that, on average, we could harness 90\% of the profit made by the best expert. 
\bibliography{template}

\end{document}